\documentclass[a4paper,aps,prd,onecolumn,preprintnumbers,showpacs,nofootinbib]{revtex4}
\usepackage{amsmath,amssymb,graphics,epsfig,subfigure}
\usepackage{color}

\begin{document}

\title{Weak cosmic censorship conjecture in Kerr black holes of modified gravity}

\author{Bin Liang \footnote{liangb2016@lzu.edu.cn},
        Shao-Wen Wei \footnote{weishw@lzu.edu.cn, corresponding author},
        Yu-Xiao Liu \footnote{liuyx@lzu.edu.cn, corresponding author}}
\affiliation{Center for Gravitational Physics, Lanzhou University, Lanzhou 730000, China\\Institute of Theoretical Physics, Lanzhou University, Lanzhou 730000, China}

\begin{abstract}
By neglecting the effects of self-force and radiation, we investigate the possibility of destroying the Kerr-MOG black hole through the point particle absorption process. Using the instability of event horizon and equation of particle motion, we get the upper and lower energy bounds allowed for a matter particle to produce the naked singularity. We find that the energy gap always exists between the upper and lower energy bounds for both extremal and near-extremal black holes, which means some tailored particles can actually lead to the violation of the weak cosmic censorship conjecture. However, when considering the effect of the adiabatic process, the result shows that the Kerr-MOG black hole gets more stable instead of a naked singularity, and thus the weak cosmic censorship conjecture can be restored at some level.
\end{abstract}


%\keywords{}

\maketitle

\section{Introduction}
\label{1}

It is widely acknowledged that spacetime singularity will be formed at the end of the gravitational collapse of regular matter~\cite{Hawking}. And all physics we have known will be invalid at the singularity. But spacetime singularity cannot be seen by distant observers, which means that it is always hidden inside the event horizon of a black hole~\cite{Penrose}. This situation is guaranteed by the weak cosmic censorship conjecture (WCCC). It is the foundation of black hole physics, especially the theorems of black hole thermodynamics. Although WCCC is believed to be correct at some level, but the test of its validity is still a huge challenge in astrophysics. For example, it was shown in Ref.~\cite{Joshi} that naked singularity can be produced during the collapse of a massive matter cloud with regular initial data.

The first work of testing the validity of WCCC can be traced back to the gedanken experiment proposed by Wald~\cite{Wald}. In the case, he envisaged that a matter particle with various quantities such as the mass, electric charge, and angular momentum is dropped into an extremal Kerr-Newman (KN) black hole. Result indicates that such particle which can destroy the horizon of the black hole either miss, or is repelled by the black hole. Later, the test of WCCC was revisited by many authors. Hubeny showed that it is indeed possible to overcharge a near-extremal Reissner-Nordstrom black hole by tossing a charged test particle, which obviously leads to the violation~\cite{Hubeny}. Also, Jacobson and Sotiriou found that a near-extremal Kerr black hole can be overspun by a test particle with tailored angular momentum~\cite{Jacobson}. But once considering the effects of radiation and self-force, it is inspiring that Kerr black holes are prohibited to be overspun so that WCCC can retain. Radiative effects occur for some special orbits but self-force always makes effects during the process that particles fall into the black holes~\cite{Barausse,Barausse2,Colleoni,Colleoni2,Hod,Zimmerman}. Neglecting the effects of self-force and radiation, different methods were proposed to destroy different black holes. For example, the extremal charged black hole was first considered to be overspun by a spinning particle in Ref.~\cite{Felice}. Studies on spinning up the higher dimensional black holes and AdS black holes were first carried out, respectively, in Refs.~\cite{Lopez,Rocha2}. Interestingly, the authored found that the quantum tunneling process can lead to the violation of WCCC~\cite{Matsas}. There are also many other works discussing about the validity of WCCC in various black hole backgrounds~\cite{Siahaan,Gwak,Gao,Gwak2,Rocha,Crisford,Gwak3,Husain,Revelar,An,Ge,
Fairoos,Hod2,Gao2,Crisford2,Natario,Duztas,Gwak4,Sorce,Yu,Saa}. However, it is worthwhile to note that due to the narrow allowed range of particle's energy to destroy the black holes, taking the radiative and self-force effects into account may be a cure for the problem that thought experiments break down WCCC. In other situations, many works have studied the possibility of destroying the horizon of a black hole through fields rather than the matter particles~\cite{Duztas2,Duztas3,Duztas4,Duztas5,Casals,Matsas2}. Although most fields can encounter the phenomenon of superradiance which might prevent the violation of WCCC, there are still some fields like massless fermions that do not exhibit superradiance~\cite{Unruh}. Thus, event horizon can be destructed if the fields are absorbed by the black hole~\cite{Saa2}.

On the other hand, among various modified gravitational (MOG) theories, the scalar-tensor-vector gravity~\cite{Moffat} has gained an increasing attention in recent years. It can provide better interpretation for the astronomical observations, such as the solar system observations~\cite{Rahvar,Toth}, the dynamics of galactic clusters~\cite{Brownstein,Moffat2}, as well as the detection of gravitational waves~\cite{MoffatMoffat,Moffat5}. Recently, the black hole solution was obtained by solving the field equations in this gravity and some novel results were revealed~\cite{Moffat3}. The shape of the shadow for the black hole was investigated in Ref.~\cite{Moffat4}. Thermodynamical properties for the non-rotating and rotating black hole solutions were studied in~\cite{Mureika,Pradhan}. The property of the accretion for the black hole was examined in Refs.~\cite{John,Armengol}. And in Refs.~\cite{Lee,Sharif,Zakria}, the geodesics were explored.

In this paper, by neglecting the effects of self-force and radiation, we use the thoughts of Wald's gedanken experiment to investigate the possibility of destroying the Kerr-MOG black holes. We find that the test particle within narrow range of energy can produce the naked singularity from both extremal and near-extremal black hole. Thus, it implies that the validity of WCCC can be broken down. Next we introduce an adiabatic process that assumes the absorption of test particles happens during a period of time rather than at a moment. It is inspiring that the violation can be reduced and the Kerr-MOG black holes remain stable during the particle absorption process. So it can be concluded that WCCC is still valid for Kerr-MOG black holes once considering the adiabatic process.

The paper is organized as follows. In Sec.~\ref{2}, we briefly review the Kerr-MOG black hole. In Sec.~\ref{3}, we get the upper and lower bounds of energy needed by test particles to destroy black hole and find that both extremal and near-extremal Kerr-MOG black holes can be overspun. In Sec.~\ref{4}, we introduce the adiabatic process with particle absorption and restore WCCC. Finally, the discussions and conclusions are presented in Sec.~\ref{5}.

\section{Kerr-MOG black hole}
\label{2}

Here we would like to give a brief review of the modified gravity. Its action contains four parts, the GR (tensor) part $S_{GR}$, the vector part $S_{\phi}$, the scalar part $S_{S}$, and the matter part, which is given by\cite{Moffat}
\begin{eqnarray}
 S&=&S_{GR}+S_{\phi}+S_{S}+S_{M},\\
 S_{GR}&=&\frac{1}{16\pi}\int d^{4}x\sqrt{-g}\frac{R}{G},\\
 S_{\phi}&=&\int d^{4}x\sqrt{-g}\left(-\frac{1}{4}B^{\mu\nu}B_{\mu\nu}
     +\frac{1}{2}\mu^{2}\phi^{\mu}\phi_{\mu}\right),\\
 S_{S}&=&\int d^{4}x\sqrt{-g}\frac{1}{G^{3}}
  \left(\frac{1}{2}g^{\mu\nu}\nabla_{\mu}G\nabla_{\nu}G-V(G)\right)
  +\int d^{4}x\frac{1}{\mu^{2}G}
  \left(\frac{1}{2}g^{\mu\nu}\nabla_{\mu}\mu\nabla_{\nu}\mu-V(\mu)\right).
\end{eqnarray}
Note that $G$ here is not a constant but a scalar field depending on the spacetime coordinates. Proca type massive vector field $\phi^{\mu}$ has mass $\mu$. And $V(G)$ and $V(\mu)$ are two potentials, which are corresponded to scalar fields $G(x)$ and $\mu(x)$. Moreover, the tensor field $B_{\mu\nu}$ are constructed with the Proca type massive vector field $\phi^{\mu}$ as $B_{\mu\nu}=\partial_{\mu}\phi_{\nu}-\partial_{\nu}\phi_{\mu}$. In addition, $B_{\mu\nu}$ satisfies 
\begin{eqnarray}
 &&\nabla_{\nu}B^{\mu\nu}=0,\\
 &&\nabla_{\sigma}B_{\mu\nu}+\nabla_{\mu}B_{\nu\sigma}+\nabla_{\nu}B_{\sigma\mu}=0.
\end{eqnarray}
For the vector field, the energy momentum tensor reads
\begin{eqnarray}
 T_{\phi\mu\nu}&=&-\frac{1}{4\pi}\left(B_{\mu}^{\;\sigma}B_{\nu\sigma}
      -\frac{1}{4}g_{\mu\nu}B^{\sigma\beta}B_{\sigma\beta}\right).
\end{eqnarray}
It was shown in Ref.~\cite{Toth} that the effect of the mass $\mu$ just becomes obvious at kiloparsec scales from the source. So if only considering the strong gravity effect of the black hole, we can neglect it on seeking a black hole solution. Then a vacuum black hole solution is described by the following action
\begin{eqnarray}
 S=\int d^{4}x\sqrt{-g}\left(\frac{R}{16\pi G}-\frac{1}{4}B^{\mu\nu}B_{\mu\nu}\right).
\end{eqnarray}
For simplicity, one can consider $G$ as a constant independent of the spacetime coordinates, while it can vary freely. Thus, the corresponding field equation is
\begin{eqnarray}
 G_{\mu\nu}=-8\pi G T_{\phi\mu\nu}.\label{field}
\end{eqnarray}
For the sake of clarity, the relation between the parameter $G$ and Newton's gravitational constant $G_{N}$ can be expressed as
\begin{eqnarray}
 G=G_{N}(1+\alpha).\label{gg}
\end{eqnarray}
It is clear that when $\alpha=0$, this gravity will reduce to GR. So $\alpha$ can be treated as a deviation parameter of MOG from GR. Employing Eq. (\ref{gg}), the gravitation charge of vector field $\phi_{\mu}$ is $\sqrt{\alpha G_{N}}m$, which implies that the parameter $\alpha$ is positive.

Solving the fields equation, one can obtain the black hole solutions. In Boyer-Lindquist coordinates, the metric of a Kerr-MOG black hole is given as~\cite{Moffat3}
\begin{eqnarray}
 ds^{2}=-\frac{\Delta}{\rho^{2}}(dt-a\sin^{2}\theta d\phi)^{2}
        +\frac{\rho^{2}}{\Delta}dr^{2}+\rho^{2}d\theta^{2}
        +\frac{\sin^{2}\theta}{\rho^{2}}(adt-(r^{2}+a^{2})d\phi)^{2}.
\end{eqnarray}
where the metric functions are given by
\begin{eqnarray}
 \Delta&=&r^{2}-2G_{N}(1+\alpha)mr+a^{2}+m^{2}G_{N}^{2}\alpha(1+\alpha),\\
 \rho^{2}&=&r^{2}+a^{2}\cos^{2}\theta.
\end{eqnarray}
In the following, we will adopt $G_{N}=1$ for simplicity. According to Ref.~\cite{Sheoran}, the ADM mass is $M=(1+\alpha)m$ and the angular momentum is $J=Ma$. Thus, we get
\begin{eqnarray}
 \Delta=r^{2}-2Mr+a^{2}+\frac{M^{2}\alpha}{(1+\alpha)}.
\end{eqnarray}
By solving $\Delta=0$, we can obtain the radius of the black hole horizon
\begin{eqnarray}
 r_{\pm}=M\pm\frac{\sqrt{(1+\alpha)(M^{2}-a^{2}(1+\alpha))}}{1+\alpha}.\label{rh}
\end{eqnarray}
It is clear that such black hole can possess two horizons for $M^{2}>a^{2}(1+\alpha)$, one degenerate horizon for $M^{2}=a^{2}(1+\alpha)$, and no horizon related to naked singularity for $M^{2}<a^{2}(1+\alpha)$. When $a=0$, this black hole becomes a static one, which also has two horizons located at
\begin{eqnarray}
 r_{\pm}=M(1\pm\frac{1}{\sqrt{1+\alpha}}),
\end{eqnarray}
where one radius is larger than $M$, while another one is smaller than it. Further setting $\alpha=0$, it will back to the Schwarzschild one with only one horizon located at $r=2M$. 

From above, one can find that this Kerr-MOG black hole is very similar to the KN black hole in GR, which is also a vacuum solution, while with an electric field. However, there are two main differences. Firstly, they are from two different gravity theory. The parameter $\alpha$ measures the deviation of the gravitational constant in the MOG from the Newton one. While the charge of the KN black hole is a hair of the black hole, and has no relation with the Newton's gravitational constant. Secondly, they have different thermodynamics. For example, the KN black hole obeys the Beckenstein-Hawking entropy-area law that the entropy equals a quarter of the black hole's horizon area. However, in the MOG, this entropy-area law will not be held. And it will be modified, see Ref.~\cite{Mureika}. So these two black hole solutions are different from each other. In the following, we will examine its geodesics.

In order to describe the process that a test particle is absorbed by the Kerr-MOG black hole, we necessarily show the geodesics of the test particle with unit mass just in terms of the radial and $\theta$-directional momentum~\cite{Moffat3}
\begin{eqnarray}
 \rho^{2}\dot{r}&=&\sigma_{r}\sqrt{\mathcal{R}},\label{rmotion}\\
 \rho^{2}\dot{\theta}&=&\sigma_{\theta}\sqrt{\Theta},
\end{eqnarray}
where
\begin{eqnarray}
 \mathcal{R}&=&(-E(r^{2}+a^{2})+aL)^{2}-\Delta(\mathcal{K}+\mu^{2}r^{2}),\\
 \Theta&=&\mathcal{K}-a^{2}\mu^{2}\cos^{2}\theta-(aE\sin^{2}\theta-L)^{2}\csc^{2}\theta.
\end{eqnarray}
The sign functions $\sigma_{r}=\pm$ and $\sigma_{\theta}=\pm$ are independent from each other, $\mathcal{K}$ is the carter constant, and $\mu^{2}$=1 and 0 for massive particle and photon, respectively. Then, by removing the separate constant $\mathcal{K}$, one gains the energy formula describing a future-forwarding particle when it passes through the event horizon of the black hole along the equator plane with $\theta=\frac{\pi}{2}$:
\begin{eqnarray}
 E_{h}=\frac{aL+r_{h}^{2}\left|\dot{r}\right|}{a^{2}+r_{h}^{2}}.\label{eee}
\end{eqnarray}
Due to the conserved quantities such as energy and angular momentum of the test particle, the Kerr-MOG black hole will undergo infinitesimal changes constrained by this relation.

\section{Violation of weak cosmic censorship conjecture}
\label{3}

Next, by neglecting the effects of radiation and self-force, we will investigate whether it is possible to destroy the horizon of the Kerr-MOG black hole. First, we introduce a parameter
\begin{eqnarray}
 \delta=M^{2}-a^{2}(1+\alpha),\label{ini}
\end{eqnarray}
In order to treat the matter particle as a test body, we assume that the energy and angular momentum of the particle satisfy $E_{h}=dM$, $L=dJ$. Therefore, according to (\ref{eee}), when the test particle reaches the event horizon, the final change of the black hole energy should be constrained as
\begin{eqnarray}
 dM=\frac{a dJ+r_{h}^{2}\left|\dot{r}\right|}{a^{2}+r_{h}^{2}}.\label{dM}
\end{eqnarray}
It is natural that the violation of WCCC should be related to the particle absorption with certain energy and angular momentum. Therefore, the test particle needs to reach the event horizon of Kerr-MOG black hole from the outside. Based on this assumption, we can easily obtain the lower energy bound by requiring $\left|\dot{r}\right|\geq 0$,
\begin{eqnarray}
 dM\geq E_{min}=\frac{a dJ}{a^{2}+r_{h}^{2}}.\label{emin}
\end{eqnarray}

On the other hand, particle absorption infinitesimally changes the corresponding charges of the black hole, so overall parameter change might cause the violation of WCCC which will produce the naked singularity. The relation of parameter is given below
\begin{eqnarray}
 &&M'^{2}-a'^{2}(1+\alpha)<0,\label{aab}
\end{eqnarray}
where the parameters $M'$ and $a'$ are
\begin{eqnarray}
 M'&=&M+dM,\\
 a'&=&\frac{J+dJ}{M+dM}. \label{inequ}
\end{eqnarray}
By solving the inequality (\ref{aab}), we can gain the upper energy bound
\begin{eqnarray}
 dM<E_{max}=-M+\left(\frac{M^{2}(dJ+J)^{2}(M^{2}-\delta)}{J^{2}}\right)^{\frac{1}{4}}.\label{emax}
\end{eqnarray}

Next we will examine whether Kerr-MOG black hole can be destroyed through the test particle absorption process.

\subsection{Extremal Kerr-MOG black hole}

As discussed above, extremal Kerr-MOG black hole corresponds to $\delta$=0. For this case, the lower and upper energy bounds read
\begin{eqnarray}
 E_{min}&=&\frac{ J MdJ}{J^2+M^4},\\
 E_{max}&=&-M+\left(\frac{M^{4}(dJ+J)^{2}}{J^{2}}\right)^{\frac{1}{4}}.
\end{eqnarray}
For the angular momentum $dJ$=0, one easily obtains
\begin{eqnarray}
 E_{min}=E_{max}=0.
\end{eqnarray}
Thus, the weak cosmic censorship holds for absorbing a test particle with vanished angular momentum, i.e., $dJ$=0.

\begin{figure}
\center{\subfigure[{$\alpha$=0}]{\label{1.1}
\includegraphics[width=7cm]{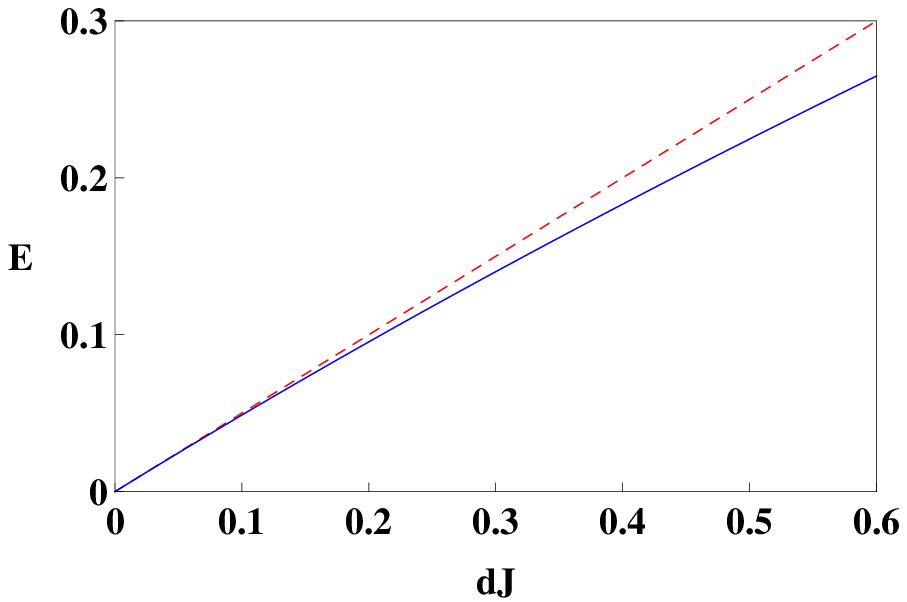}}
\subfigure[{$\alpha$=0.3}]{\label{1.2}
\includegraphics[width=7cm]{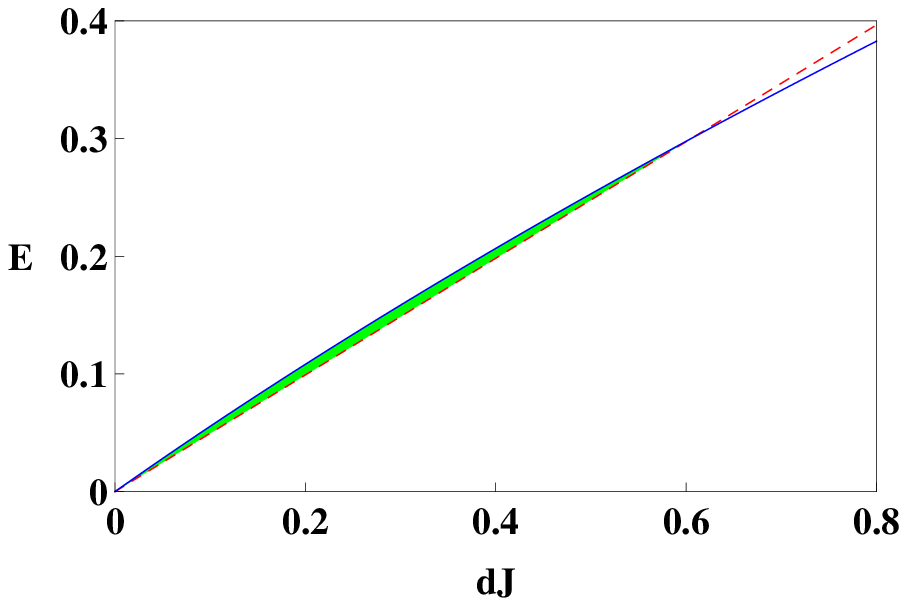}}
\subfigure[{$\alpha$=0.8}]{\label{1.3}
\includegraphics[width=7cm]{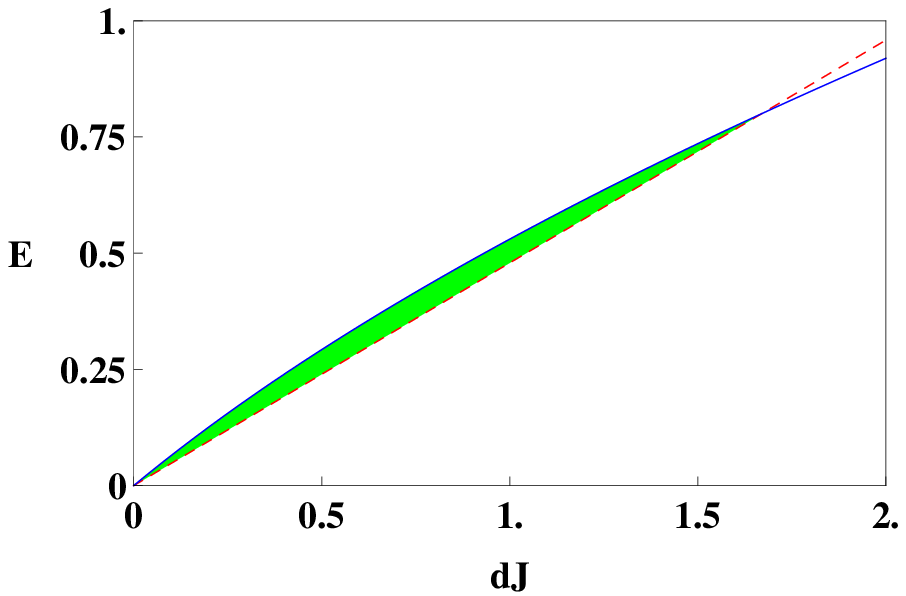}}
\subfigure[{$\alpha$=1.5}]{\label{1.4}
\includegraphics[width=7cm]{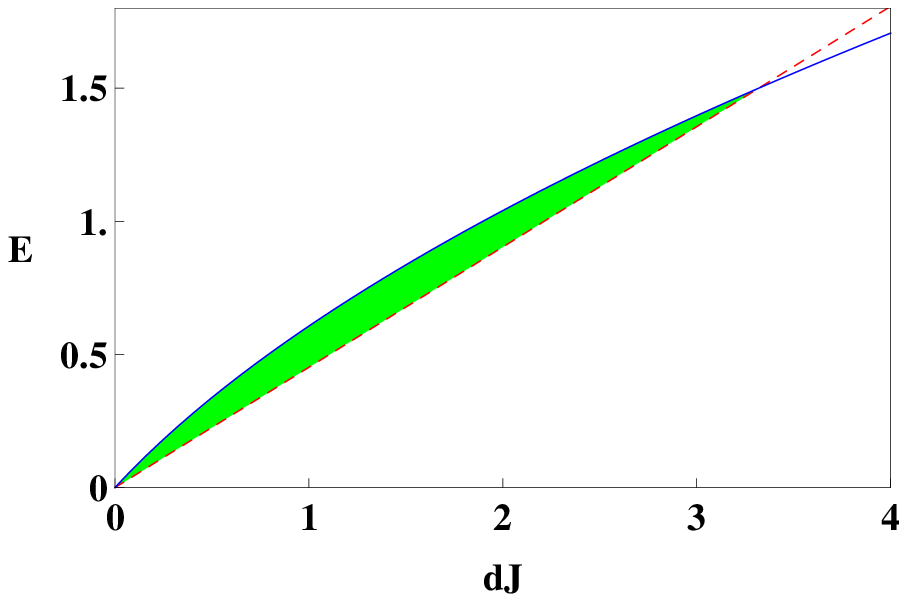}}}
\caption{Energy bounds $E_{max}$ (blue solid line) and $E_{min}$ (red dashed line) vs. $dJ$ for the extremal Kerr-MOG black hole with $M$=1. (a) $\alpha$=0. (b) $\alpha$=0.3. (c) $\alpha$=0.8. (d) $\alpha$=1.5. The green-dark region is for $E_{max}>E_{min}$.}\label{fig1}
\end{figure}

On the other hand, for $dJ\neq 0$, we plot the energy bounds as a function of the angular momentum $dJ$ in Fig.~\ref{fig1} for different values of $\alpha$. For $\alpha=0$, the Kerr black hole will be recovered. As is shown in Fig.~\ref{1.1}, one can clearly have $E_{min}>E_{max}$, which implies that extremal Kerr black hole cannot be overspun by a test particle, and this result is consistent with that obtained by Wald~\cite{Wald}. However, when $\alpha\neq0$, the result becomes interesting. From Fig.~\ref{fig1}, it is clearly that, for small $dJ$, $E_{max}>E_{min}$, therefore the extremal Kerr-MOG black hole can become a naked singularity by absorbing a test particle of small angular momentum. While for large $dJ$, the situation will be reversed and one will have $E_{min}>E_{max}$. Thus the black hole will not be destroyed. Moreover, with the increase of the deformed parameter $\alpha$, the region allowed to destroy the event horizon enlarges, which can be found by comparing with Figs.~\ref{1.2},~\ref{1.3}, and~\ref{1.4}.

%%%%%%%%%%%%%%%%%%%%%%%%%%%%%%%%%%%%%%%%%%%%%%%%%%%%%%%%%%%%%%%%%%%%%
%%%%%%%%%%%%%%%%%%%%%%%%%%%%%%%%%%%%%%%%%%%%%%%%%%%%%%%%%%%%%%%%%%%%%%%%%

\subsection{Near extremal Kerr-MOG black hole}

\begin{figure}
\center{\subfigure[{$dJ$=0.01}]{\label{2.1}
\includegraphics[width=7cm]{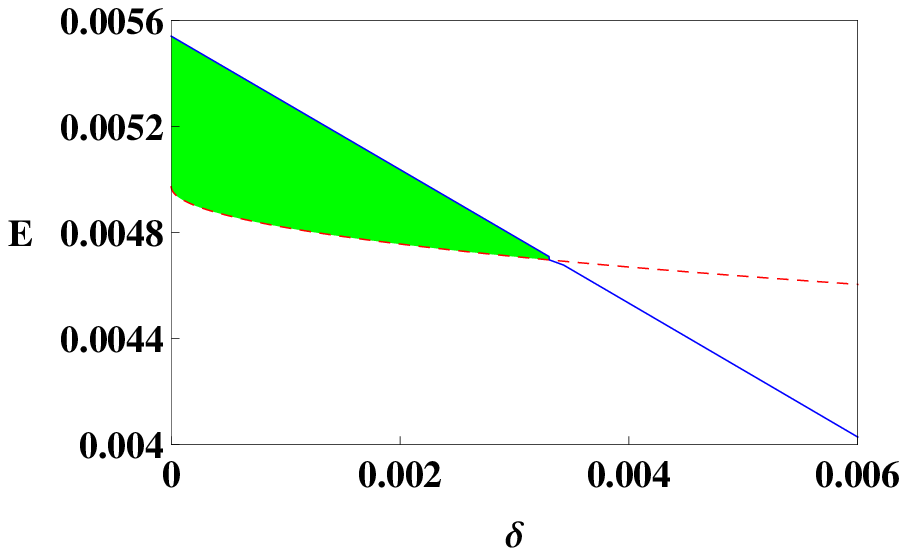}}
\subfigure[{$dJ$=0.1}]{\label{2.2}
\includegraphics[width=7cm]{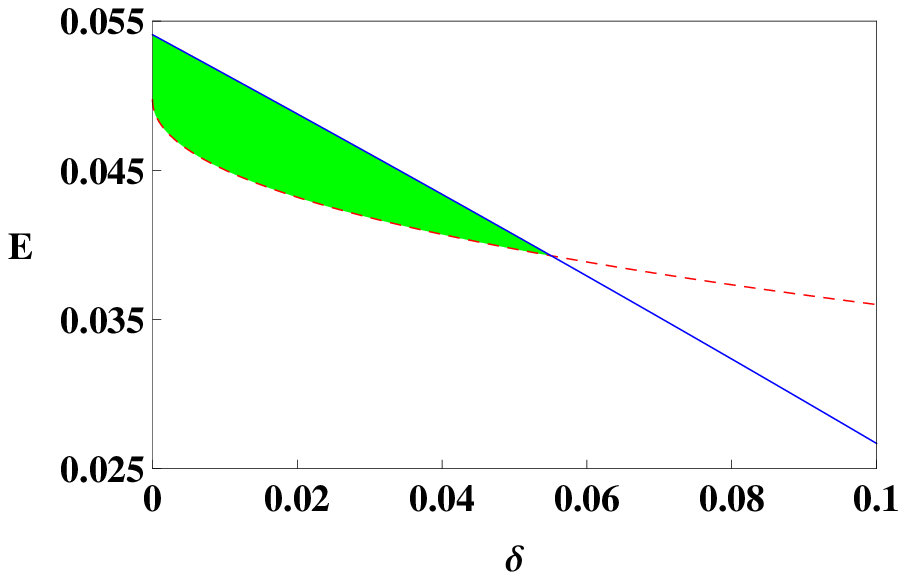}}}
\caption{Energy bounds $E_{max}$ (blue solid line) and $E_{min}$ (red dashed line) vs. $\delta$ with $M$=1 and $J$=0.9. (a) $dJ$=0.01. (b) $dJ$=0.1. The green-dark region is for $E_{max}>E_{min}$.}\label{fig2}
\end{figure}

For the near-extremal Kerr-MOG black hole, its satisfies $0\le\delta\ll 1$. By taking $M$=1 and $J$=0.9, we plot the lower and upper energy bounds as a function of $\delta$ for the test particle with angular momentum $dJ$=0.1 and 0.01 in Fig.~\ref{fig2}. Both the bounds decrease with $\delta$. In Fig.~\ref{fig2}, we can find that there always exists a region (denoted with dark green color) which satisfies $E_{max}>E_{min}$. Therefore, the near-extremal Kerr-MOG black hole can also be destroyed like the extremal Kerr-MOG black hole. It is also shown that, with the increase of $\delta$, the energy gap decreases and ultimately vanishes at some point beyond which the near-extremal black hole will not be destroyed by absorbing a test particle. Moreover, for smaller angular momentum $dJ$, the energy region allowed to violate WCCC shrinks.

%%%%%%%%%%%%%%%%%%%%%%%%%%%%%%%%%%%%%%%%%%%%%%%%%%%%%%%%%%%%%%%%%%%%%
%%%%%%%%%%%%%%%%%%%%%%%%%%%%%%%%%%%%%%%%%%%%%%%%%%%%%%%%%%%%%%%%%%%%%%%%%

\begin{figure}
\center{\subfigure[{$dJ$=0.01}]{\label{3.1}
\includegraphics[width=7cm]{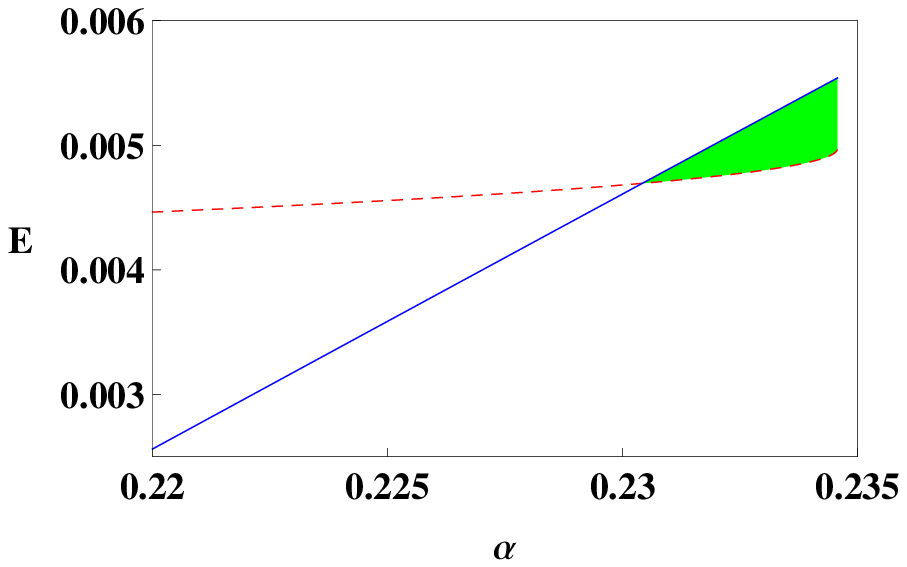}}
\subfigure[{$dJ$=0.1}]{\label{3.2}
\includegraphics[width=7cm]{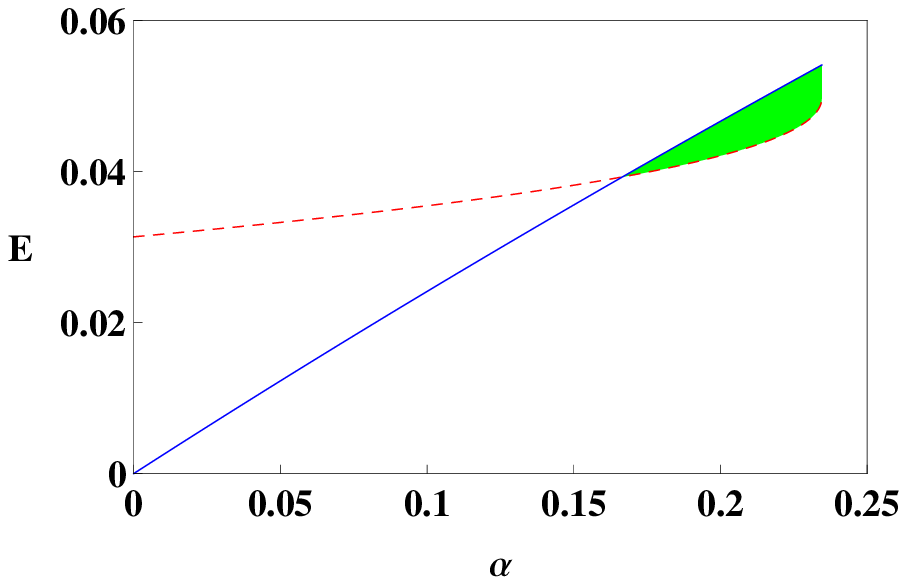}}}
\caption{Energy bounds $E_{max}$ (blue solid line) and $E_{min}$ (red dashed line) vs. $\alpha$ with $M$=1 and $J$=0.9. (a) $dJ$=0.01. (b) $dJ$=0.1. The green-dark region is for $E_{max}>E_{min}$. Extremal black hole is located at $\alpha=0.2346$.}\label{fig3}
\end{figure}

In Fig.~\ref{fig3}, we plot the lower and upper upper energy bounds as a function of the deformed parameter $\alpha$ with $M=1$ and $J=0.9$ for $dJ$=0.1 and 0.01, respectively. For this case, the extremal Kerr-MOG black hole corresponds to $\alpha=0.2346$. So we require $\alpha<0.2346$ for a black hole. From the figure, it is easy to find that both the bounds increase with $\alpha$. For small $\alpha$, we have $E_{max}<E_{min}$, so these black holes are not allowed to be destroyed. However, when $\alpha$ is larger than some certain values near the extremal value, for example, $\alpha$=0.2304 for $dJ$=0.01 and $\alpha$=0.1666 for $dJ$=0.1, one will get $E_{max}>E_{min}$. This indicates that the near extremal black hole can be overspun by absorbing the particle. On the other hand, we can also find that the parameter range of $\alpha$ increases with $dJ$.

%%%%%%%%%%%%%%%%%%%%%%%%%%%%%%%%%%%%%%%%%%%%%%%%%%%%%%%%%%%%%%%%%%%%%
%%%%%%%%%%%%%%%%%%%%%%%%%%%%%%%%%%%%%%%%%%%%%%%%%%%

Based on the discussion above, WCCC can be violated both for the extremal and near-extremal Kerr-MOG black holes. In particular, the greater deformed parameter $\alpha$ is, the larger energy gap to destroy the event horizon becomes. Thus it can be concluded that the existence of deformed parameter $\alpha$ increases the possibility to destroy a rotating Kerr black hole.

\section{Restoration of weak cosmic censorship conjecture}
\label{4}

As suggested in Refs.~\cite{Gwak,Chirco}, then we would like to consider a continuous path called the adiabatic process during the particle absorption. The violation of WCCC is mostly due to the fact that the absorption of the test particle is assumed to happen at a moment, while it should take time to spread across the horizon of the black hole. Instead of jumping from the initial condition to the final state at a once, we split the process into $N$ steps.

When the black hole parameters change, respectively, from $M$ and $J$ to $M+dM$ and $J+dJ$, the parameter $\delta$ will change to $\delta'$
\begin{eqnarray}
 \delta'=-\Big(\frac{J+dJ}{M+dM}\Big)^2(1+\alpha)+(M+dM)^2. \label{delta'}
\end{eqnarray}
For simplicity, we define $F(M, J, \delta, dJ)=\delta'$. Then substituting Eqs. (\ref{rh}), (\ref{ini}), and (\ref{dM}) into the above equation \eqref{delta'}, we can clearly obtain
\begin{eqnarray}
 F(M,J,\delta,dJ)&=&-\frac{(dJ+J)^2 M^4(M^2-\delta) \left[M^5-M^3\delta+J^2\left(M+2\sqrt{\frac{M^2(M^2-\delta)\delta}{J^2}}\right)\right]^2}
 {J^2\left[dJ J M(M^2-\delta)+M^4(M+\left|\dot{r}\right|)(M^2-\delta)+J^2\Big(M^3+\left|\dot{r}\right|\delta
 +2M(M+\left|\dot{r}\right|)\sqrt{\frac{M^2(M^2-\delta)\delta}{J^2}}\Big)\right]^2}\nonumber\\
 &+&\Bigg[M
          +\frac{dJ J M(M^2-\delta)+\left|\dot{r}\right|\left(M^6+J^2\delta-M^4\delta
                 +2J^2 M\sqrt{\frac{M^2(M^2-\delta)\delta}{J^2}}\right)}
           {M\left(M^5-M^3\delta+J^2\left(M+2\sqrt{\frac{M^2(M^2-\delta)\delta}{J^2}}\right)\right)}\Bigg]^2.
\end{eqnarray}
Here, $\left|\dot{r}\right|$ is the radial velocity of the test particle. We can also understand this case as that a particle of energy $dM$ and angular momentum $dJ$ is split into $N$ small ones with each one has energy $dM/N$ and angular momentum $dJ/N$. Then these small particles are absorbed by black hole one by one.  Then the continuous path also known as the adiabatic process is given in terms of $N$ steps
\begin{eqnarray}
 \delta&&\rightarrow F(M,J,\delta,\frac{dJ}{N})=\delta_{1}\rightarrow F\left(M+\frac{dM}{N},J+\frac{dJ}{N},\delta_{1},\frac{dJ}{N}\right)=\delta_{2}
 %\rightarrow F(M+2\frac{dM}{N},J+2\frac{dJ}{N},\delta_{2},\frac{dJ}{N})=\delta_{3}
 \nonumber\\
 &&\rightarrow ~\cdots ~\rightarrow F\left(M+(N-1)\frac{dM}{N},J+(N-1)\frac{dJ}{N},\delta_{N-1},\frac{dJ}{N}\right)=\delta_{N}.
\end{eqnarray}
In each of the steps, it is natural to assume the radial velocity $\left|\dot{r}\right|$ keeps unchanged. The final state of the black hole is still $M+dM$ and $J+dJ$, but the resulting parameter $\delta_{N}$ can always be positive if the step number $N$ is taken to be large enough. Thus, the event horizon will not be destroyed, which leads to the restoration of WCCC. In Table.~\ref{tab}, we list the changes of small parameter $\delta$, the horizon radius $r_h$ and the deformed parameter $\alpha$. The numerical values which we choose are $M=1$, $J=0.99$, and $dJ=0.01$. In addition, the initial value and the step number is $\delta=0$ and $N=10$. According to the relation (\ref{dM}), we set $\left|\dot{r}\right|=0.00005$ and $dM=0.005025$. It is well noticed that, if the value of radial velocity $\left|\dot{r}\right|$ vanishes, then the step number needs to be infinity which cannot be shown through numerical calculation.

Instead of destroying the event horizon of the extremal Kerr-MOG black hole at once, the black hole gets further away from the extremal condition with the step $N$. Moreover, we find that the horizon radius becomes larger within this adiabatic process, which may mean that the black hole gets more stable during the absorption of the test particle rather than produces the naked singularity. What's most interesting is that deformed parameter $\alpha$ decreases during the process which indicates Kerr-MOG black holes gradually approach to more stable Kerr black holes in accordance with the above discussion. This process also applies to near-extremal Kerr-MOG black hole. Thus, we restore the validity of WCCC when considering effect of the adiabatic process.

%%%%%%%%%%%%%%%%%%%%%%%%%%%%%%%%%%%%%%%%%%%%%%%%%%%%%%%%%%%%%%%%%%%%%%%%%%%%
\begin{table}[h]
\begin{center}
\begin{tabular}{cccccccccccc}
  \hline\hline
  $N$ & 0 & 1 & 2 & 3 & 4 & 5 & 6 & 7 & 8 & 9 & 10\\\hline
  $\delta$ $(10^{-4})$ & 0.00 & 0.81 & 1.46 & 2.04 & 2.59 & 3.10 & 3.59 & 4.06 & 4.50 & 4.92 & 5.33\\
  $r_h$ & 1.0000 & 1.0094 & 1.0130 & 1.0157 & 1.0179 & 1.0200 & 1.0218 & 1.0235 & 1.0250 & 1.0265 & 1.0279\\
  $\alpha$ $(10^{-2})$ & 2.030 & 2.021 & 2.014 & 2.007 & 2.001 & 1.995 & 1.990 & 1.985 & 1.980 & 1.975 & 1.971\\\hline\hline
\end{tabular}
\caption{Changes of $\delta$, $r_h$ and $\alpha$ with $N$.}\label{tab}
\end{center}
\end{table}
%%%%%%%%%%%%%%%%%%%%%%%%%%%%%%%%%%%%%%%%%%%%%%%%%%%%%%%%%%%%%%%%%%%%%%%%%%%%%%

\section{Discussions and Conclusions}
\label{5}

In this paper, we investigated whether the validity of WCCC still remains in Kerr-MOG black hole background without taking self-force and radiation into consideration. By using the equation of motion of a test particle with energy and angular momentum, we got the infinitesimal change of the corresponding charge of the black hole during the absorption process of the test particle.

Next, we studied the instability of the event horizon and the constraint condition that the matter particle can actually reach the event horizon during the absorption process. Thus, we gained the upper and lower energy bounds needed by a test particle to destroy the Kerr-MOG black hole. It is found that there always exists a narrow energy range allowed to destroy the event horizon for both extremal and near-extremal black holes. Besides, with larger deformed parameter $\alpha$, the energy gap for test particle to destroy Kerr-MOG black hole becomes greater which is a major difference from GR. We concluded that the existence of the deformed parameter $\alpha$ makes the possible violation of WCCC for extremal Kerr black holes.

Further, we also considered the effect of the adiabatic process. Instead of assuming the absorption process happens at a moment, the whole process is split into many steps during a period of time. In this case, the Kerr-MOG black hole becomes larger and more stable rather than a naked singularity further confirmed by the fact that deformed parameter $\alpha$ decreases during this process. Thus, WCCC is restored. In conclusion, the WCCC can be violated for the extremal and near extremal Kerr-MOG black hole by absorbing a test particle. While when considering the adiabatic process in the test particle absorption process, the validity of WCCC can be guaranteed.

\section*{Acknowledgements}
This work was supported by the National Natural Science Foundation of China (Grants No. 11675064, No. 11522541, No. 11375075, and  No 11205074), and the Fundamental Research Funds for the Central Universities (Grants Nos. lzujbky-2016-115).

\end{document}